\documentclass[aps,prd,twocolumn,showpacs,amsfonts]{revtex4}
\usepackage{amsmath}
\usepackage{graphicx}

\begin{document}

\title{\bf Braneworld cosmological models with anisotropy}

\author{Antonio Campos$^{1,2}$, Roy Maartens$^2$, David Matravers$^2$,
and Carlos F. Sopuerta$^2$}

\address{~}

\address{$^1$Institut f\"ur Theoretische Physik,
             Universit\"at Heidelberg,
             D-69120~Heidelberg,
             Germany}

\address{$^2$Institute of Cosmology and Gravitation, University of
Portsmouth, Portsmouth~PO1~2EG, UK}

\date{\today}

\begin{abstract}

For a cosmological Randall-Sundrum braneworld with anisotropy,
i.e., of Bianchi type, the modified Einstein equations on the
brane include components of the five-dimensional Weyl tensor for
which there are no evolution equations on the brane. If the bulk
field equations are not solved, this Weyl term remains unknown,
and many previous studies have simply prescribed it ad hoc. We
construct a family of Bianchi braneworlds with anisotropy by
solving the five-dimensional field equations in the bulk. We
analyze the cosmological dynamics on the brane, including the Weyl
term, and shed light on the relation between anisotropy on the
brane and Weyl curvature in the bulk. In these models, it is not
possible to achieve geometric anisotropy for a perfect fluid or
scalar field -- the junction conditions require anisotropic stress
on the brane. But the solutions can isotropize and approach a
Friedmann brane in an anti-de Sitter bulk.

\end{abstract}


\maketitle

\section{Introduction}

High-energy physics theories have recently inspired relatively
simple phenomenological models in which one can test some of the
consequences of string theories. Randall and
Sundrum~\cite{Randall:1999ee,Randall:1999vf} proposed a model that
captures some of the essential features of the dimensional
reduction of eleven-dimensional supergravity introduced by
Ho$\check{\mbox{r}}$ava and
Witten~\cite{Horava:1996qa,Horava:1996ma}. The second
Randall-Sundrum (RS2) scenario~\cite{Randall:1999vf} is a
five-dimensional Anti-de Sitter ($AdS_5$) ``bulk" spacetime with
an embedded Minkowski 3-brane where matter fields are confined and
Newtonian gravity is effectively reproduced at low energies. The
RS2 scenario was generalized to a Friedmann-Robertson-Walker (FRW)
brane, showing that the Friedmann equation at high energies gives
$H^2\sim\rho^2$, in contrast with the general-relativistic
behaviour $H^2\sim
\rho$~\cite{Binetruy:1999ut,Binetruy:1999hy,Csaki:1999jh,
Cline:1999ts}.

As shown in~\cite{Shiromizu:1999wj}, the modified field equations
on the brane have two new contributions from extra-dimensional
gravity:
\begin{equation}\label{modfe}
G_{\alpha\beta}=-\Lambda g_{\alpha\beta}+ \kappa^2 T_{\alpha\beta}
+ 6{\kappa^2 \over \lambda}{\cal S}_{\alpha\beta} - {\cal
E}_{\alpha\beta}\,,
\end{equation}
where $\lambda$ is the brane tension (the vacuum energy of the
brane when $T_{\alpha\beta}=0$), and $\Lambda$ and $\kappa$ are
the four-dimensional cosmological and gravitational constants
respectively, given in terms of $\lambda$ and the fundamental
constants of the bulk $(\Lambda_5\,,\kappa_5)$ by:
\begin{equation}\label{con}
\Lambda = {\Lambda_5 \over 2} + {\lambda^2 \over
12}\kappa_5^2\,,~\kappa^2= {\lambda \over 6}\kappa_5^4 \,.
\end{equation}
The term ${\cal S}_{\alpha\beta}$ is quadratic in
$T_{\alpha\beta}$ and dominates at high energies ($\rho>\lambda$).
The five-dimensional Weyl tensor is felt on the brane via its
projection, ${\cal E}_{\alpha\beta}$. This Weyl term is determined
by the bulk metric, not by equations on the brane. In FRW
braneworlds, the bulk is
Schwarzschild-$AdS_5$~\cite{Bowcock:2000cq,
Kraus:1999it,Ida:1999ui}, and ${\cal E}_{\alpha\beta}$ reduces to
a simple Coulomb term that gives rise to ``dark radiation" on the
brane. The simplest generalizations of FRW braneworlds are Bianchi
braneworlds.

By making an assumption about the Weyl term on the brane, the
dynamics of a Bianchi type~I brane were studied
in~\cite{Maartens:2000az}, and it was shown that high-energy
effects from extra-dimensional gravity remove the anisotropic
behaviour near the singularity that is found in general
relativity. This was extended via a phase space analysis of
Bianchi~I and V braneworlds~\cite{Campos:2001pa,Campos:2001cn},
showing that the anisotropy is negligible close to the singularity
for perfect fluid models with a barotropic linear equation of
state $p=w\rho$, with $w\geq 0$ a constant, opposite to the
general relativity case. It was then suggested that this may be
generic in cosmological braneworlds, which was supported by
subsequent work~\cite{Coley:2001ab,Coley:2001va} (see
also~\cite{Hervik:2002im,vandenHoogen:2002hb,Coley:2003ua}).
However, a perturbative analysis~\cite{Bruni:2002bv} suggests that
this may only be true for homogeneous models.

These studies, and others~\cite{Santos:2001vr,Toporensky:2001hi,
Chen:2001wg,Chen:2001sm,Paul:2001ce,Barrow:2001pi,Savchenko:2002hx,
Paul:2002md,Harko:2002hv,vandenHoogen:2002hc,Aguirregabiria:2003at,
Aguirregabiria:2003yd}, considered only the dynamical equations on
the brane, making various assumptions about the Weyl term in the
absence of knowledge of the bulk metric. In~\cite{Frolov:2001wz} a
bulk metric with a Kasner brane was presented. However, since the
Kasner metric is a solution of the 4-dimensional Einstein vacuum
equations, the bulk metric is a simple warped extension; the
general result, with the generic form of the bulk metric, is given
in~\cite{Brecher:1999xf}~\footnote{Note that this result is
sensitive to the form of the bulk field equations, and it breaks
down in the presence of a Gauss-Bonnet term in the gravitational
action~\cite{Barcelo:2002wz}.}. The simplest example of this
general result is a Minkowski brane, leading to the RS2 solution.
Another example is the Schwarzschild black string
solution~\cite{Chamblin:1999by}. Up to now, no complete solutions,
i.e., for the brane and bulk metrics, have been found for
cosmological Bianchi
braneworlds~\footnote{In~\cite{Cadeau:2000tj}, solutions with an
anisotropic bulk containing a black hole with a non-spherical
horizon were found.}. The key difficulty is to find anisotropic
generalizations of $AdS_5$ that can incorporate anisotropy on a
cosmological brane, and that are necessarily non-conformally flat.

Previous studies of Bianchi braneworlds have considered the
effects of ${\cal S}_{\alpha\beta}$, under various assumptions on
${\cal E}_{\alpha\beta}$. Here we tackle the question of the
construction of complete models for cosmological braneworlds with
anisotropy, that is, we want to construct both the metric in the
bulk and on the 3-brane, so that ${\cal E}_{\alpha\beta}$ is
determined and not assumed ad hoc.

\section{The geometry of the bulk}

For a five-dimensional bulk spacetime with a negative cosmological
constant, $\Lambda_5< 0$, and no additional matter sources, the
Einstein equations are:
\begin{equation}
^{5\!}G_{AB}+\Lambda_5\, ^{5\!}g_{AB}= 0\,. \label{efe}
\end{equation}
In order to construct cosmological braneworlds with anisotropy we
start from the ansatz used
by~\cite{Binetruy:1999ut,Binetruy:1999hy} (see also
\cite{Mukohyama:1999qx,Vollick:1999uz}):
\begin{eqnarray}
^{5\!}ds^2 = -n^2(t,y) dt^2  + a^2(t,y) d\Sigma^2_k + b^2(t,y)
dy^2 \,, \label{bdl}
\end{eqnarray}
where $d\Sigma^2_k$ is the line element of the three-dimensional
maximally symmetric surfaces $\{t=t_\ast,y=y_\ast\}$, with
curvature curvature index $k=0,\pm 1$. Clearly, all the
hypersurfaces $\{y=y_\ast\}$ inherit a FRW metric. Although the
Einstein equations~(\ref{efe}) can be completely solved for the
metric (\ref{bdl}), the explicit complete solution (bulk+brane)
(see~\cite{Binetruy:1999ut,Binetruy:1999hy}) was found for
$\dot{b}=0$, which corresponds to Gaussian normal coordinates
adapted to the foliation with normal $n_Adx^A = dy$. Since the
bulk is Schwarzschild-$AdS_5$, an alternative approach is based on
a moving brane in static spherical bulk
coordinates~\cite{Kraus:1999it,Ida:1999ui},
\begin{equation}
^{5\!}ds^2 = -f(r) dT^2 + \frac{dr^2}{f(r)} + r^2 d\Sigma^2_k \,,
\label{Sch-AdS}
\end{equation}
where
\begin{equation}
f(r) = k + \frac{r^2}{\ell^2} - \frac{\mu}{r^2} \,. \label{fder}
\end{equation}
Here $\ell^2=-6/\Lambda_5$, so that $\ell$ is the curvature scale
of the bulk. When the parameter $\mu$, the mass of the bulk black
hole, vanishes, the solution is simply $AdS_5$, so that the bulk
Weyl tensor, and hence the brane Weyl term, vanish. If $\mu\neq
0$, then the tidal field of the black hole generates a non-zero
Weyl term on the brane. The existence of the black hole horizon
requires that $\mu\geq0$ for a flat or closed geometry, and
$\mu\geq-\ell^2/4$ for the open case. The brane trajectory is
$r=a(\tau)$, where $\tau$ is cosmological proper time on the brane
and $a(\tau)$ is the scale factor, whose evolution is determined
by the junction conditions. For a $Z_2$-symmetric brane, this
gives the modified Friedmann equation on the brane,
\begin{equation}
H^2+{k \over a^2} = {\kappa^2 \over 3} \rho\left(1+ {\rho \over
2\lambda} \right)+ {\Lambda \over 3} + {\mu \over a^4}\,,
\label{friedmanneq}
\end{equation}
where the high-energy correction term is $\rho^2/\lambda$ and the
last term on the right is the dark radiation term.

A natural extension of the ansatz in Eq.~(\ref{bdl}) that will
introduce anisotropy is (compare~\cite{Cadeau:2000tj} for a
similar approach):
\begin{equation}
^{5\!}ds^2 = -n^2(t,y) dt^2 + h_{IJ}(t,y)\omega^I\omega^J +
b(t,y)^2dy^2 \,,
\end{equation}
where $\omega^I = \omega^I_i dx^i$ are 1-forms invariant under a
Bianchi group (see~\cite{Stephani:2003tm} for details), and
$h_{IJ}\omega^I\omega^J$ is the metric induced on the surfaces
$\{t=t_\ast\,,y=y_\ast\}$. For simplicity, we consider only
Bianchi type~I models ($\omega^I_i=\delta^I_i$) (the procedure for
non-Abelian Bianchi groups is essentially the same) with a
diagonal metric $h_{IJ}$:
\begin{equation}
^{5\!}ds^2 = -a^2_0(t,y)dt^2 + \sum_{i=1}^3 a^2_i(t,y) (dx^i)^2+
b(t,y)^2dy^2 \,. \label{gbi}
\end{equation}
The field equations~(\ref{efe}) for this metric are non-linear
partial differential equations (PDEs) in $(t,y)$, like the field
equations for the metric~(\ref{bdl}).  In the case of the metric
(\ref{bdl}) the $\{ty\}$-component of the field equations provides
a relation that leads to a set of first integrals. However, this
procedure does not work for Eq.~(\ref{gbi}), and one must deal
with non-linear PDEs. We have not been able to find a procedure to
solve them analytically.

These difficulties indicate that in order to find analytic
solutions we should abandon the generic case and consider special
solutions that do not require PDEs. We try a static and Gaussian
normal ansatz,
\begin{equation}
^{5\!}ds^2  = - e^{2A_0(y)}dt^2+ \sum^{3}_{i=1}e^{2A_i(y)}(dx^i)^2
+ dy^2\,, \label{metric}
\end{equation}
where we pay the price that the brane is no longer static in the
coordinate system. This ansatz can in fact be seen as a
five-dimensional generalization of a similar
ansatz~\cite{Linet:1986sr} used in the search for four-dimensional
static and cylindrically symmetric spacetimes describing cosmic
strings in the presence of a non-vanishing cosmological constant.
The field equations for the metric~(\ref{metric}) are
\begin{eqnarray}
A''_\alpha +  A'_\alpha \sum^{3}_{\beta=0} A'_\beta -
\frac{\omega^2}{4} & =&  0  \,,  \label{efe1}\\
\sum_{0\leq\alpha<\beta\leq 3} \!\!\! A'_\alpha A'_\beta -
\frac{3}{8}\omega^2 &=&  0\,, \label{efe2}
\end{eqnarray}
where $\omega=4/\ell$.

In order to solve these equations, we introduce the determinant of
the metric,
\begin{equation}
u^2(y) = \exp\left(2\sum_{\alpha=0}^3 A_\alpha\right)\,. \label{udef}
\end{equation}
Multiplying Eq.~(\ref{efe1}) by $u(y)$ and summing over $\alpha$,
we get
\begin{equation}
  u''-\omega^2 u =  0\,,
\end{equation}
with first integral,
\begin{equation}
  u'^2 - \omega^2 u^2 + C =  0\,,
\label{eq:integral}
\end{equation}
where $C$ is an arbitrary constant of integration. Once $u(y)$ is
known, $A_\alpha(y)$ can be obtained by quadrature:
\begin{equation}
A'_\alpha =  \frac{1}{4}\frac{u'}{u}+ \frac{C_\alpha}{u}\,,
\label{eq:A'}
\end{equation}
which comes from the integration of Eq.~(\ref{efe1}). The
constants $C_\alpha$ are constrained by Eqs.~(\ref{efe2}) and
(\ref{udef}):
\begin{eqnarray}
\sum^{3}_{\alpha=0} C_\alpha   & = & 0\,, \label{con1} \\
\sum_{0\leq \alpha < \beta \leq 3} C_\alpha C_\beta  & = &
\frac{3}{8} C\,. \label{con2}
\end{eqnarray}
These imply
\begin{equation}
(C_1+C_2)^2+(C_1+C_3)^2+(C_2+C_3)^2 = -\frac{3}{4} C \,.
\label{constr}
\end{equation}
Taking the square of Eq.~(\ref{con1}) and using Eq.~(\ref{con2})
yields equivalently
\begin{equation}
\sum_{\alpha=0}^3 C^2_\alpha = -\frac{3}{4}C \,. \label{constr2}
\end{equation}
Thus $C$ can never be positive, and the $C_\alpha$'s must be the
coordinates of a three-sphere of radius $\sqrt{-3C}/2$, hence
$|C_\alpha|\leq \sqrt{-3C}/2$.

When $C=0$ the parameters $C_\alpha$ must all be zero as well. In
this particular case,
\begin{equation}
A_\alpha(y) =  A^o_\alpha + \frac{\omega}{4}(y-y_o)\,,
\end{equation}
where $A^o_\alpha$ are integration constants.  This model
corresponds, as expected, to an exact $AdS_5$ bulk.

Thus, we are left to consider negative values for $C$, and we
rewrite it as $C = -\omega^2 B^2$. By Eq.~(\ref{eq:integral}),
\begin{equation}
u(y) = B \sinh \left[\omega (y-y_o)\right]\,, \label{eq:u_sol}
\end{equation}
and then Eq.~(\ref{eq:A'}) gives
\begin{equation}
A_\alpha(y) =  A^o_\alpha + \frac{1}{4}\ln \left|u(y)\right| +
C_\alpha v(y)\,,
\end{equation}
where $v'=1/u $, so that
\begin{equation}
v(y) =  \frac{1}{2\omega B} \ln \left\{ \frac{\cosh \left[\omega
(y-y_o)\right]-1} {\cosh \left[\omega (y-y_o)\right]+1} \right\}
\,.
\end{equation}

Finally, we can write the bulk metric solution as
\begin{eqnarray}
e^{2A_\alpha} &=& N^2_\alpha \big|\sinh \left[\omega
(y-y_o)\right] \big|^{1/2}  \nonumber\\ &&~{} \times \left\{
\frac{\cosh \left[\omega (y-y_o)\right]-1} {\cosh \left[\omega
(y-y_o)\right]+1} \right\}^{q_\alpha}\,, \label{met_coef}
\end{eqnarray}
where $q_\alpha=C_\alpha/\omega B$ (recall that $\omega B\neq 0$),
and $N_\alpha$ are constants whose value can be chosen by
rescaling coordinates, but which satisfy the constraint
\begin{equation}
\prod_{\alpha=0}^3 N^2_\alpha = B^2 \,,
\end{equation}
which follows from Eqs.~(\ref{udef}) and (\ref{eq:u_sol}). The
exponents $q_\alpha$ are constrained by Eqs.~(\ref{con1}) and
(\ref{constr2}):
\begin{equation}
\left| q_\alpha\right|~ \leq~ \frac{3}{4}\,.
\end{equation}
Note that this is a more restrictive bound than the one found
above only from Eq.~(\ref{constr2}).

We consider first the special case $C_1=C_2=C_3=-C_0/3$, with two
possible sets of parameters in Eq.~(\ref{met_coef}), namely
$(q^\pm_0,q^\pm_i) =(\mp {3\over4}, \pm {1\over4})$. These two
special cases are Schwarzschild-$AdS_5$, with $k=0$, written in
Gaussian normal coordinates. (The $k=-1$ case corresponds to
Bianchi~V, and the $k=1$ case to Bianchi~IX.) We see this via a
coordinate transformation in the metric of Eq.~(\ref{Sch-AdS}):
\begin{equation}
r^4 = \frac{8\mu}{\omega^2}(1\mp\cosh \left[\omega
(y-y_o)\right])\,, \label{coord_trans}
\end{equation}
and the remaining coordinates are rescaled by constants that
depend on $\mu$, $\omega$, and $N_\alpha$. It follows that
$q^+_\alpha$ leads to a negative mass $\mu$, which we exclude, so
that $q^-_\alpha$ is the physical solution (with a black hole
horizon).

This shows that our general five-dimensional bulk solution,
Eq.~(\ref{met_coef}), can be seen as an {\em anisotropic
generalisation of Schwarzschild-$AdS_5$}. This distinguishes our
anisotropic solution from the vacuum Kasner
braneworld~\cite{Frolov:2001wz}.

We now investigate the character of the singular point $y=y_o$ via
curvature scalars. It turns out to be more convenient to use a new
set of constants,
\begin{eqnarray}
\frac{d_1}{2} = q_2+q_3\,,~ \frac{d_2}{2} = q_1+q_3\,,~
\frac{d_3}{2} = q_1+q_2\,,
\end{eqnarray}
with
\begin{equation}
d^2_1+d^2_2+d^2_3=3\,. \label{const_for_d}
\end{equation}
The isotropic cases $q^\pm_\alpha$ correspond to the points $(\pm
1, \pm 1, \pm 1)$ on the 2-sphere~(\ref{const_for_d}). The square
of the bulk Weyl tensor, ${\cal C}^2 = \,^{5\!}C_{ABCD}\,
^{5\!}C^{ABCD}$ is given by
\begin{eqnarray}
{\cal C}^2  &=& \frac{\omega^4}{16\sinh^4[\omega(y-y_o)]}
\left\{21 -\left(d^4_1+d^4_2+d^4_3 \right)\right.\nonumber\\ &&
\left.~{}+ 18\cosh^2[\omega(y-y_o)] \right.\nonumber\\ &&
\left.~{} +36 d_1d_2d_3\cosh[\omega(y-y_o)] \right\} \,.
\end{eqnarray}
The behaviour near $y_o$ is
\begin{equation}
{\cal C}^2 \underset{y\rightarrow y_o}{\longrightarrow}~
\frac{\left[ 39 +36\, d_1d_2d_3 -\left(d^4_1+d^4_2+d^4_3\right)
\right]}{16(y-y_o)^4}\,.
\end{equation}
There are four sets of constants $d_i$ that lead to zero
numerator: $D_1 = (-1,1,1)\,,$ $D_2 = (1,-1,1)\,,$ $D_3 =
(1,1,-1)\,,$ and $D_4 = (-1,-1,-1)$. For these cases, ${\cal C}^2$
is regular at $y_o$:
\begin{eqnarray}
{\cal C}^2 = \frac{9\,\omega^4}{8\left\{\cosh[\omega(y-y_o)]+1
\right\}^4} \,.
\end{eqnarray}
For all other values of the $d_i$'s, $y=y_o$ is a curvature
singularity. The cases $D_1$, $D_2$, and $D_3$ are equivalent in
the sense that they represent the same spacetime. The case $D_4$
is Schwarzschild-$AdS_5$ (with positive mass), and $y=y_o$
corresponds to the horizon of the black hole, since
$f(r(y=y_o))=0$, and not to the singularity. Thus the Gaussian
coordinates only cover the exterior of the black hole.

Far from $y=y_o$, ${\cal C}^2$ decays exponentially,
\begin{equation}
{\cal C}^2 \underset{y\rightarrow \infty}{\longrightarrow}~
\frac{9}{2}\,\omega^4\, e^{-2\omega y}\,.
\end{equation}
This behaviour, which is completely independent of the parameters
$d_i$ (or $C_i$), means that our general anisotropic solution is
asymptotically $AdS_5$. The square of the bulk Riemann tensor, the
Kretschmann scalar ${\cal R}^2 = \,^{5\!}R_{ABCD}\,
^{5\!}R^{ABCD}$, is
\begin{equation}
{\cal R}^2 =  {\cal C}^2 +\frac{5}{32}\,\omega^4\,.
\end{equation}

\section{Embedding of the brane}

In order to obtain Bianchi~I cosmological models the embedding of
the brane must respect the Bianchi~I symmetries, so the most
general embedding is
\begin{equation}
t=S(\tau)\,,~~ x^i=X^i\,,~~ y=Y(\tau)\,, \label{embed}
\end{equation}
where $\{\tau,X^i\}$ are local coordinates on the brane.  The
normal to the brane is
\begin{equation}
n_A dx^A = \frac{\epsilon}{\sqrt{1-V ^2}}\left( -V  e^{A_0} dt +
dy \right)\,. \label{normal}
\end{equation}
Here $\epsilon = \pm 1$ determines the orientation of the normal,
and $V  $ is a function defined by the coordinate velocity of the
brane,
\begin{equation}
V ^2 = \frac{\dot{Y}^2}{1+\dot{Y}^2}\,, \label{sigma}
\end{equation}
so that $|V| \leq 1$.  The functions $S$ and $Y$ are not
independent; since $n_Adx^A$ must vanish identically on the brane,
\begin{equation}
\dot{S}^2 = (1+\dot{Y}^2)e^{-2A_0(Y)}\,. \label{norma}
\end{equation}

We use a local foliation of the bulk such that the brane is itself
a hypersurface of the foliation. This foliation is described by
the normal~(\ref{normal}), with $V $ being now a function of $y$.
The brane is then determined by the choice~(\ref{embed})
and~(\ref{sigma}). An alternative way of determining the location
of the brane, which will be also useful, is to prescribe the
function $V (y)$ and then the embedding is given by
Eqs.~(\ref{sigma}) and (\ref{norma}) up to an integration
constant.

We introduce the vectors
\begin{eqnarray}
u_A dx^A &=& \frac{1}{\sqrt{1-V ^2}} \left( -e^{A_0}dt + V  dy
\right) \,, \label{timevec}\\ \mbox{e}^i_A dx^A &=& e^{A_i} dx^i
\,,
\end{eqnarray}
which together with $n_A$ form an orthonormal basis for the bulk.
The vector ${u}^A$ is a four-velocity tangent to the foliation,
and hence to the brane. The condition~(\ref{norma}) ensures that
$\tau$ is the proper time on the brane of the observers with
four-velocity $u^A$.

The metric inherited by the brane and other hypersurfaces of the
foliation, is the first fundamental form,
\begin{equation}
g_{AB}= \,^{5\!}g_{AB}-n_An_B\,,
\end{equation}
so that
\begin{eqnarray}
&& g_{tt} = -\frac{e^{2A_0}}{1-V ^2}\,,~~~~ g_{ij} =
e^{2A_i}\;\delta_{ij} \,, \label{indm1} \\
&& g_{ty} = -V  e^{-A_0} g_{tt}\,, ~~~~ g_{yy} = V ^2 e^{-2A_0}
g_{tt}\,. \label{indm2}
\end{eqnarray}
The extrinsic curvature (second fundamental form) is
\begin{equation}
K_{AB}=  \frac{1}{2}\pounds^{}_{n} g_{AB} = g_{A}^Cg_{B}^D\,
\,^{5\!}\nabla^{}_Cn^{}_D \,,
\end{equation}
where $\pounds $ is the Lie derivative and $K_{AB}=K_{(AB)}$,
$K_{AB}n^B = 0$. Then
\begin{eqnarray}
K_{tt} &=& -\epsilon\frac{e^{2A_0}}{(1-V ^2)^{3/2}} \left\{ A'_0
+\frac{VV'}{(1-V ^2)}\right\}\!, \label{kAB1} \\
K_{ij} &=& \epsilon \frac{e^{2A_i}}{\sqrt{1-V ^2}}
A'_i\;\delta_{ij} \,,
\\
K_{ty} &=& -V  e^{-A_0} K_{tt}\,,\\ K_{yy} &=& V ^2 e^{-2A_0}
K_{tt}\,, \label{kAB2}
\end{eqnarray}
with trace
\begin{equation}
K = \frac{\epsilon}{\sqrt{1-V ^2}}\left\{ \frac{u'}{u}
+\frac{VV'}{(1-V ^2)} \right\}\,.
\end{equation}
Using Eq.~(\ref{sigma}), one can give a geometrical interpretation
to terms in the extrinsic curvature. The factor $\sqrt{1-V ^2}$ is
$\sqrt{1+\dot{Y}^2}$, i.e., the inverse of the arc-length of the
embedding function $Y(\tau)$, whereas the $VV'$ term can be
written as ${\ddot{Y}}/{(1+\dot{Y}^2)}$, i.e., the curvature times
the arc-length.

\subsection{Projection of the bulk Weyl tensor onto the
brane}

The modified Einstein equations~(\ref{modfe}) on the brane contain
the projection of the bulk Weyl tensor~\cite{Shiromizu:1999wj},
\begin{equation}
{\cal E}_{AC} = \,^{5\!}C_{ABCD}n^Bn^D\,,
\end{equation}
which is symmetric, tracefree and orthogonal to $n^A$. Relative to
any observer, and in particular the observer with the preferred
four-velocity $u^A$, this can be decomposed
as~\cite{Maartens:2000fg,Maartens:2001jx}
\begin{equation}
{\cal E}_{A B} = -\kappa^2 \left\{ {\cal U}\left(u_A
u_B+\frac{1}{3}h_{AB}\right) + 2{\cal Q}_{(A}u_{B)} + {\cal
P}_{AB} \right\}\,,
\end{equation}
where $h_{AB}=g_{AB}+u_A u_B$ projects into the comoving rest
space of $u^A$, ${\cal U}$ is the Weyl energy density on the
brane, ${\cal Q}_A$ is the Weyl momentum flux on the brane, and
${\cal P}_{AB}$ is the Weyl anisotropic stress on the brane.

Bianchi-I symmetry enforces ${\cal Q}_{A}=0$, while
\begin{eqnarray}
{\cal U} &=& \frac{1}{2\kappa^2 u^2}\left\{
C_0\left(u'-2C_0\right) + \frac{3}{8}\omega^2B^2\right\}\,,\\
{\cal P}_{AB} &=& {1 \over \kappa^2} \sum^3_{i=1} {\cal
P}_i\,\mbox{e}_{iA}\mbox{e}_{iB}\,,~~ \sum^3_{i=1} {\cal P}_i =
0\,,
\end{eqnarray}
where
\begin{eqnarray}
{\cal P}_i &=&
\frac{C_0+3C_i}{3u}\left(\frac{u'}{4u}+\frac{C_0}{u}\right)
\nonumber\\&&~{} -\frac{1}{u(1-V ^2)}\left[(C_0+3C_i)
\left(\frac{u'}{4u}+\frac{C_i}{u}\right) \right. \nonumber\\
 &&~{}\left. +\frac{\omega^2B^2 - 16C^2_i}{4u} \right] \,.
\label{bpi}
\end{eqnarray}
Clearly, ${\cal P}_{AB} = 0$ for the isotropic case.

\section{Braneworld matter fields}

While the induced metric is continuous, there are discontinuities
in its first derivatives across the brane, so that there is a jump
in the extrinsic curvature. In the case of ${Z}_2$-symmetry with
the brane as fixed point, the junction conditions determine the
brane energy-momentum tensor in terms of the extrinsic curvature:
\begin{equation}
{T}_{AB} - \lambda g_{AB} =- \frac{2}{\kappa^2_5}\left( K_{AB}-K
g_{AB}\right) \,. \label{ctab}
\end{equation}

The energy-momentum tensor can be decomposed, relative to
observers with four-velocity $u^A$, as
\begin{equation}
{T}_{A B} = \rho u_A u_B + ph_{A B} +2q_{(A}u_{B)}+\pi_{AB}\,,
\end{equation}
where $\rho$, $p$, $q_A$, and $\pi_{AB}$ are, respectively, the
energy density, isotropic pressure, momentum density and
anisotropic stresses measured by $u^A$.

For a Bianchi~I braneworld, the symmetries enforce $q_A=0$. From
Eqs.~(\ref{kAB1})-(\ref{kAB2}) and (\ref{ctab}), we find that for
our Bianchi~I models,
\begin{eqnarray}
\rho +\lambda  & = & \frac{2\epsilon}{\kappa^2_5\sqrt{1-V ^2}}
\left( \frac{C_0}{u} -\frac{3}{4}\frac{u'}{u} \right)\,,
\label{rho}\\
p - \lambda &=& \frac{2\epsilon}{\kappa^2_5\sqrt{1-V ^2}} \left\{
\frac{3}{4}\frac{u'}{u} +\frac{1}{3}\frac{C_0}{u}+
\frac{VV'}{\left(1-V ^2\right)} \right\}\!. \label{p}\\
\pi_{AB} &=&  \sum^3_{i=1}
{\pi}_i\,\mbox{e}_{iA}\mbox{e}_{iB}\,,~~ \sum^3_{i=1} \pi_i = 0\,,
\label{piij}
\end{eqnarray}
where
\begin{eqnarray}
\pi_i = -\frac{2\epsilon}{\kappa^2_5\sqrt{1-V ^2}}
\left(\frac{C_0+3C_i}{3u} \right) \,. \label{pii}
\end{eqnarray}
The anisotropic directional pressures $p_i = p + \pi_i$ are
\begin{eqnarray}
p_i -\lambda = \frac{2\epsilon}{\kappa^2_5\sqrt{1-V ^2}} \left\{
\frac{3}{4}\frac{u'}{u}-\frac{C_i}{u}+ \frac{VV'}{\left(1-V
^2\right)} \right\}\!.
\end{eqnarray}

In our case, since we do not have momentum density, the vanishing
of the anisotropic stresses implies a perfect-fluid
energy-momentum tensor. This happens if and only if $C_0=-3C_i$,
for all $i$; i.e., {\em the brane can support a perfect fluid if
and only if the metric is isotropic}.  Furthermore,
Eqs.~(\ref{bpi}) and (\ref{pii}) show that {\em the Weyl
anisotropic stresses ${\cal P}_i$ vanish if and only if the matter
anisotropic stresses $\pi_i$ vanish.} Therefore, geometric
anisotropy enforces, via the extrinsic curvature and the junction
conditions, anisotropy in the matter fields. This may be a
peculiar feature of our solution, based on the ansatz
Eq.~(\ref{metric}). However, it may be a generic feature of
anisotropic cosmological braneworlds.

The fluid kinematics of the matter are described by the expansion,
$\Theta = \nabla_A u^A$, the shear, $\sigma_{AB} = [h_{(A}^C
h_{B)}^D-{1 \over 3}h^{CD}h_{AB}] \nabla_Cu_D$, the vorticity,
$\omega_{AB} = h_{[A}^Ch_{B]}^D \nabla_Du_C$, and the
acceleration, $\dot{u}^A = u^B\nabla_B u^A\,.$ For Bianchi
symmetry, the matter flow is geodesic and irrotational,
$\omega_{AB}=0= \dot{u}_A$. The expansion and shear for our
Bianchi~I braneworlds are given by
\begin{eqnarray}
\Theta &=& \frac{V }{\sqrt{1-V ^2}}\left(\frac{3}{4}\frac{u'}{u}
-\frac{C_0}{u} \right)\,,\label{theta} \\ \sigma_{AB} &=&
\sum^3_{i=1} \sigma_i\, \mbox{e}_{iA}\mbox{e}_{iB}\,,~~~
\sum^3_{i=1} \sigma_i = 0\,, \label{sigmaab}
\end{eqnarray}
where
\begin{equation}
\sigma_i = \frac{V }{\sqrt{1-V ^2}}\left(
\frac{C_0+3C_i}{3u}\right). \label{sigmai}
\end{equation}

Equations~(\ref{rho}) and (\ref{theta}) imply
\begin{equation}
\rho+\lambda =-{2\epsilon \over \kappa_5^2 V}\,\Theta\,.
\end{equation}
To ensure that the brane is expanding for positive energy density,
we require $\epsilon/V<0$. Equations~(\ref{pii}) and
(\ref{sigmai}) also imply
\begin{equation}
\pi_{AB}=-{2\epsilon \over \kappa_5^2 V}\,\sigma_{AB}\,.
\end{equation}

We have checked that our expressions satisfy the generalized
Friedmann equation for a Bianchi
brane~\cite{Maartens:2000az,Maartens:2000fg}:
\begin{eqnarray}
{1\over 9}\Theta^2 &=& {\kappa^2 \over 3} \rho\left(1+ {\rho \over
2\lambda} \right)-\frac{\kappa^2}{4 \lambda}\pi^{AB}\pi_{AB} +
{\Lambda \over 3}\nonumber\\ &&~{}+ {\kappa^2 \over 3} {\cal U}+
{1\over 6} \sigma^{AB}\sigma_{AB}\,.
\end{eqnarray}

\section{Some explicit models}

We can construct explicit cosmological models using the freedom
still available in embedding the 3-brane. Choosing the parameters
$q_\alpha$, which are subject to the constraints~(\ref{con1}) and
(\ref{con2}), defines the bulk spacetime. The embedding $Y(\tau)$
is a function of one variable (proper time), and involves the
freedom to choose the direction of the normal $n^A$, and the sign
$\epsilon$ which defines its orientation.

One way to construct a particular cosmology on the 3-brane is to
prescribe the density $\rho$. Using Eq.~(\ref{rho}), $V $ can then
be obtained as a function of $y$,
\begin{equation}
V ^2 = 1 -\left[\frac{2}{\kappa_5^2(\rho +\lambda)} \left(
\frac{C_0}{u} -\frac{3}{4}\frac{u'}{u}\right) \right]^2\,.
\end{equation}
Then the embedding is completely determined by integrating,
\begin{equation}
\tau-\tau_o  = \pm \int^y \frac{\sqrt{1-V ^2}}{V }dy,
\end{equation}
which gives an implicit form of the function $Y(\tau)$. However,
one cannot use any physical argument or intuition in order to
start with a cosmologically relevant density $\rho$ as a function
of the coordinate of the extra dimension.

A more appealing procedure is to start by prescribing the
embedding function $Y(\tau)$, or equivalently the redefined
function
\begin{equation}
x(\tau)=\omega\left[y(\tau)-y_o\right]\,.
\end{equation}
Then $V(\tau) $ and $S(\tau)$ are given by Eqs.~(\ref{sigma}) and
(\ref{norma}), respectively, and $\rho(\tau)$ by Eq.~(\ref{rho}).
Using this approach, we investigate under what circumstances a
Minkowski brane can be embedded in our anisotropic bulk, how we
can recover the standard embedding of a FRW brane in the isotropic
case, and, finally, several examples of the embedding of
anisotropic branes in a general anisotropic bulk.

\subsection{Embedding of a Minkowski brane}

The simplest embedding is $x(\tau)=x_\ast=$ const ($>0$), which
implies $V =0$, so that the $g_{AB}(\tau,X^i,x_\ast)$ are
constant, by Eqs.~(\ref{indm1}) and (\ref{indm2}), and the 3-brane
is Minkowskian. However, the matter variables are constants that
do not vanish in general,
\begin{eqnarray}
\rho_\ast & = & \lambda_c \left( \frac{3\cosh x_\ast -
4q_0}{3\sinh x_\ast} \right) -\lambda\,,\\ p_\ast & = & \lambda
-\lambda_c \left( \frac{9\cosh x_\ast + 4q_0}{9\sinh x_\ast}
\right)\,,
\end{eqnarray}
so the models obtained in this way are not empty. Here
\begin{equation}
\lambda_c = -\epsilon \frac{6}{\ell\kappa^2_5}\,, \label{lambdac}
\end{equation}
so that $|\lambda_c|$ is the critical tension corresponding to the
RS
fine-tuning~\cite{Randall:1999ee,Randall:1999vf,Shiromizu:1999wj},
i.e., for which $\Lambda=0$:
\begin{equation}
\Lambda =\frac{3}{\ell^2} \left[ \left(\frac{\lambda}{\lambda_c}
\right)^2-1 \right]\,.
\end{equation}
In general the matter fluid will not be perfect because the
anisotropic stresses~(\ref{pii}) only vanish when the bulk
spacetime is isotropic ($q_0+3q_i=0\,,$ for all $i$). Note that
taking $\epsilon=-1$ [see Eq.~(\ref{lambdac})], there always
exists a finite positive $\lambda$ such that $\rho>0$. For
instance, in the isotropic case where $q_0={3\over4}$, this
condition is satisfied by any $\lambda$ such that
\begin{equation}
\lambda < \lambda_c \sqrt{\frac{\cosh x_\ast-1}{\cosh
x_\ast+1}}\,.
\end{equation}

The brane cannot be empty: if $\rho_\ast=0=p_\ast$, $\pi_{AB}=0$,
then $q_\alpha=0$, which is incompatible with the constraint
equation~(\ref{con2}).

If we embed the brane at $x_\ast\ll1$, then
\begin{equation}
\rho_\ast \sim \lambda_c \left[ \frac{3-4q_0}{3x_\ast}
+\frac{x_\ast}{2}\right] -\lambda\,.
\end{equation}
Then, for $-{3\over4}\leq q_0 < {3\over4}$, the matter density
grows very large unless the brane tension $\lambda$ is also
unrealistically large. In the isotropic case, it decreases as
$x_\ast$ approaches the black hole horizon at $x=0$, where it
becomes negative. On the other hand, if we place the brane at a
large distance, $x_\ast\gg 1$,
\begin{equation}
\rho_\ast \sim \lambda_c-\lambda\,,~~p_\ast\sim -\rho_\ast\,.
\end{equation}
This result is independent of the parameters $q_\alpha$ so it
means we can have a nearly vacuum brane embedded in our
anisotropic bulk solution for large enough $x_\ast$ if we choose
$\lambda$ as the critical brane tension $\lambda_c$. The existence
of this embedding is something one should have expected a priori,
because our 5-dimensional solution asymptotically approaches an
$AdS_5$ spacetime for large $x$.

To sum up, we have shown that we can {\em embed a non-vacuum
Minkowskian brane in a general anisotropic bulk.} In order to make
the 3-brane empty we have to locate it asymptotically far from the
horizon. These results generalise the findings
of~\cite{Bowcock:2000cq} that a Minkowski brane can be embedded in
a (isotropic) Schwarzschild-$AdS_5$ bulk.

\subsection{Embedding of a FRW brane}

In the isotropic case, for a bulk spacetime with $q^+_\alpha$
which correspond to the point $D_4$ on the
sphere~(\ref{const_for_d}), we
follow~\cite{Kraus:1999it,Ida:1999ui} and choose
\begin{equation}
x(\tau)= \mathrm{arccosh}\left[ \frac{\omega^2}{8\mu}a^4 (\tau) -1
\right]\,.
\end{equation}
Substituting in Eq.~(\ref{rho}), we get the effective Friedmann
equation~(\ref{friedmanneq}) with $k=0$, thus recovering the
results
of~\cite{Binetruy:1999hy,Kraus:1999it,Ida:1999ui,Shiromizu:1999wj}
for an isotropic bulk. Note that in this case the anisotropic
stress tensor~(\ref{piij}) is identically zero and the matter on
the brane is a perfect fluid. A combination of
Eqs.~(\ref{rho})~and~(\ref{p}) also leads to the effective
Raychaudhuri equation \cite{Maartens:2000fg},
\begin{equation}
\dot{H} = - H^2 - \frac{\kappa^2}{6}\left(\rho+3p\right)
-\frac{\kappa^2}{6\lambda}\rho\left(2\rho+3p\right) -
\frac{\mu}{a^4} + \frac{\Lambda}{3}\,.
\end{equation}
The cosmological dynamics of this case have been extensively
investigated for a barotropic linear equation of
state~\cite{Campos:2001pa,Campos:2001cn}.

\subsection{Embedding of an anisotropic brane}

The metric tensor on the brane has Bianchi~I form,
\begin{equation}
ds^2 = -d\tau^2 + \sum_{i=1}^3 a^2_i(\tau) (dx^i)^2\,,
\end{equation}
and the mean scale factor of the universe is $a(\tau)=
(a_1a_2a_3)^{1/3}$. There are infinite ways of constructing these
models as there are infinite ways of prescribing the embedding.
Here we just present a few examples.

~

\noindent\underline{Example I}:
\begin{equation}\label{em1}
x(\tau)=\Omega\tau \,,
\end{equation}
with $0\leq \tau <\infty$ and $\Omega >0\,.$ In this case $\rho$
grows very large at early times and asymptotically reaches a
constant value at late times. Positivity of the energy density
requires that
\begin{equation}
0< \lambda \leq \lambda_c \sqrt{ 1 + \left(\frac{\Omega}
{\omega}\right)^2}\,, \label{pos_rho}
\end{equation}
where the equality corresponds to an asymptotically vacuum
universe. The anisotropic stress vanishes at late times and the
fluid becomes perfect. The universe expands exponentially in the
far future,
\begin{equation}
a(\tau) \underset{\tau\rightarrow \infty}{\longrightarrow}
e^{\Omega\tau/4}\,,
\end{equation}
independent of the constants $q_\alpha$ defining the bulk
spacetime. In the early universe,
\begin{equation}
a(\tau)   \underset{\tau\rightarrow 0}{\longrightarrow}
\tau^{(3-4{q_0})/12}\,,~~ a_i(\tau) \underset{\tau\rightarrow
0}{\longrightarrow} \tau^{(1-4q_i)/4}\,.
\end{equation}
(The exponent in $a(\tau)$ is always positive.) These cosmological
models do not isotropize as we approach the initial singularity,
in contrast with the results
of~\cite{Maartens:2000az,Campos:2001pa,Campos:2001cn}, where
assumptions were imposed on the Weyl anisotropic stresses. This
example shows that {\em the Weyl anisotropic stresses can affect
significantly the dynamical behaviour near the singularity.}

\begin{figure}[t]
\includegraphics*[totalheight=0.75\columnwidth,
width=\columnwidth]{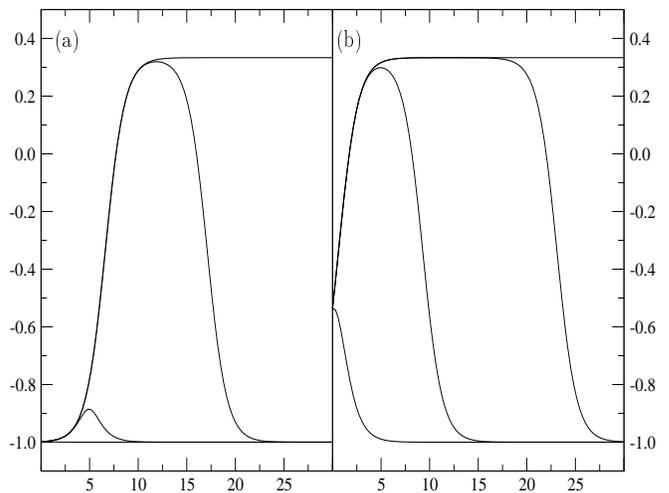} \caption{ Evolution of the
equation of state $w$ as a function of proper time $\tau/\Omega$,
Eq.~(\ref{eos}), for (a)~$q_0=-0.001$ and (b)~$q_0=-0.4$. The
curves in both graphs correspond to values $\alpha=1, 0.999999999,
0.9999, 0.1$ from top to bottom. \label{fig:eq_state}}
\end{figure}

Note also that despite the fact that the universe is collapsing in
the past, there exist models within the family of solutions for
which at least one spatial dimension could be expanding (e.g.,
$q_3=-3/4, q_0=q_1=q_2=1/4$). In the future the $a_i(\tau)$
approach the mean radius $a(\tau)$ and all the models become
isotropic. For the embedding~(\ref{em1}) the equation of state has
a simple analytical expression,
\begin{equation}\label{eos}
w(\tau)  = -\left[\frac{(1-\alpha) e^{\Omega\tau}+(1+\alpha)
e^{-\Omega\tau}+8q_0/9}{(1-\alpha)
e^{\Omega\tau}+(1+\alpha)e^{-\Omega\tau}-4q_0/3} \right],
\end{equation}
where $\alpha= \lambda/\lambda_c\sqrt{1+(\Omega/\omega)^2}$ is a
normalized brane tension, with $0<\alpha \leq1$ by
Eq.~(\ref{pos_rho}). For $q_0>0$ the equation of state becomes
singular as $\tau$ increases.   For $q_0=0$ and any value of
$\alpha$, we have $w=-1$.  As $\alpha$ approaches its maximum
value, the equation of state has a transient period with $w > 0$
before reaching the constant value $-1$. However, when $\alpha =
1$, $w$ tends to $1/3$, i.e. the matter behaves as a radiation
fluid, even though the expansion is increasing exponentially due
to geometrical effects. Some examples of the fluid behaviour
admitted by the embedding are shown in Fig.~\ref{fig:eq_state}.

~

\noindent\underline{Example II}:
\begin{equation}
x(\tau)=4\beta\ln(\Omega\tau)\,,
\end{equation}
with $\Omega^{-1}\leq \tau <\infty$ and $\Omega,\beta
>0\,.$ The qualitative behaviour is very similar to
that of Example~I. Here the brane tension has to satisfy the new
condition $0<\lambda\leq\lambda_c$ instead of Eq.~(\ref{pos_rho})
in order to have $\rho>0$. The universe isotropizes in the future,
with mean radius
\begin{equation}
a(\tau) \underset{\tau\rightarrow \infty}{\longrightarrow}
(\Omega\tau)^\beta\,,
\end{equation}
which includes radiation-domination ($\beta={1\over2}$),
matter-domination ($\beta={2\over3}$), and power-law inflation
($\beta>1$).

~

\noindent\underline{Example III}:
\begin{equation}
x(\tau)=\mathrm{arccosh}\left[\frac{(\Omega\tau)^{2\beta}+1}
{(\Omega\tau)^{2\beta}-1} \right] \,,
\end{equation}
with $\Omega^{-1}\leq \tau <\infty$ and $\Omega,\beta >0\,.$ The
scale factors are
\begin{equation}
a_i(\tau) = \left[ \frac{2(\Omega\tau)^{\beta(1-4q_i)}}
{(\Omega\tau)^{2\beta}-1} \right]^{1/4}\,,
\end{equation}
so that each spatial direction can have different rates of
expansion/contraction, and the universe does not isotropize in the
future, unlike Examples~I and II. However, the models do
isotropize in the past. The mean scale factor shows that all these
models are expanding in the past and collapsing in the future.
(In~\cite{Campos:2001cn} a similar qualitative behaviour was found
in a Bianchi~I brane when the mass of the bulk black hole is
negative.) In this case the matter content never behaves as a
perfect fluid.

\section{Discussion}

We have constructed complete (brane+bulk geometry) cosmological
braneworlds with anisotropy. These solutions are the first such
models with matter content. Our ansatz starts from a static form
for the bulk metric, Eq.~(\ref{metric}), with the brane moving
relative to the static frame. The anisotropy arises from imposing
Bianchi symmetries on a family of homogeneous 3-surfaces. For the
sake of simplicity, we only considered the Abelian Bianchi~I case,
but other groups can be treated following the same approach.

There are two important aspects of the construction of
cosmological braneworlds: \begin{itemize} \item {\em The bulk
geometry.} In our case, this is given by Eq.~(\ref{met_coef}),
where the parameters $q_\alpha$ control the anisotropy. The
anisotropic bulk curvature produces a nonzero Weyl anisotropic
tensor ${\cal P}_{AB}$ which, as shown in the examples of the
previous section, can have a fundamental impact on the dynamics.
Previous studies of Bianchi brane-world dynamics which impose ad
hoc assumptions on ${\cal P}_{AB}$ are unable to treat
consistently the relation between the bulk and brane geometries.
\item {\em The embedding.} This is where most of the freedom arises (a
function of one variable). As shown by the examples in the
previous section, the dynamics are very sensitive to the
embedding. From the physical point of view, this leads to the
question of what is the most natural state of movement for a
brane. However this question cannot be answered in the
phenomenological context of the RS2 scenarios.
\end{itemize}

In choosing the embedding of the brane it is very important to
consider the following general feature of our models:  when the
brane is close to $y=y_o$, the effects of the anisotropy are
important for the cosmological dynamics, whereas when it is
located far from $y=y_o$, we have an effectively FRW cosmological
model (in an anisotropic bulk). This fact can be seen from the
relative shear eigenvalues,
\begin{equation}
\frac{\sigma_i}{\Theta}~ \underset{x\gg1}{\longrightarrow}~
\frac{\ell}{9}\frac{C_0+3C_i}{u}~
\underset{x\rightarrow\infty}{\longrightarrow}~0 \,.
\end{equation}
This is illustrated by Examples~I and II, where in the future the
brane isotropizes.  By contrast, in Example~III the brane
approaches FRW in the past.

A striking feature of our models is that geometric anisotropy on
the brane, from the Bianchi symmetry, imposes via the bulk
curvature and the junction conditions, anisotropy on the matter
content of the brane. In other words, it is not possible within
our family of models to have a perfect fluid matter content
(including the case of a minimally coupled scalar field) --
anisotropic pressure in the matter is unavoidable unless the brane
geometry reduces to Friedmann isotropy. This feature may be a
consequence of the simplicity of the bulk metric ansatz that we
used, but it raises an interesting challenge, i.e., to find
complete Bianchi brane-world solutions with perfect fluid matter
and nonzero anisotropy.



\[ \]
{\bf Acknowledgements:}

We thank Bill Bonnor for bringing Ref.~\cite{Linet:1986sr} to our
attention. AC was supported at Portsmouth (when this work was
initiated) by a university fellowship, and is supported at
Heidelberg by the Alexander von Humboldt Stiftung/Foundation. RM
is supported by PPARC. CFS is supported by the EPSRC and thanks
the Institut f\"ur Theoretische Physik of the Universit\"at
Heidelberg for hospitality during the last stages of this work.





\end{document}